\documentclass{cs21proc}

\usepackage{kantlipsum}

\newcommand{\ha}{H$\alpha$\ }
\newcommand{\hb}{H$\beta$\ }

\editors{A. S. Brun, J. Bouvier, P. Petit}
\publisher{Zenodo}
\conference{The 21th Cambridge Workshop on Cool Stars, Stellar Systems, and the Sun}
\conferencedate{2022}

\title{Exploring the short-term variability of \ha and \hb emissions in a sample of M Dwarfs}
\author{Vipin Kumar$^{1,2,3}$,
	A. S. Rajpurohit$^{1}$,
	Mudit K. Srivastava$^{1}$}

\affiliation{$^{1}$Astronomy \& Astrophysics Division, Physical Research Laboratory, Ahmedabad 380009, India\\
	$^{2}$Department of Physics, Indian Institute of Technology, Gandhinagar, 382335, India\\
	$^{3}$I. Physikalisches Institut, Universit\"at zu K\"oln, Z\"ulpicher Stra\ss{}e 77, 50937, K\"oln, Germany}
\shorttitle{\ha variability in M-Dwarfs}
\shortauthors{Kumar et al.}

\abs{Activities in M dwarfs show spectroscopic variability over various time scales ranging from a few seconds to several hours. The time scales of such variability can be related to internal dynamics of M dwarfs like magnetic activity, energetic flaring events, their rotation periods, etc. The time variability in the strengths of prominent emission lines (particularly \ha) is mostly taken as a proxy of such dynamic behavior. In this study, we have performed the spectroscopic monitoring of 83 M dwarfs (M0-M6.5) to study the variations in \ha and \hb emissions on short-time scales. Low-resolution (resolution $\sim$5.7 angstroms) spectral time series of 3-5 minutes cadence over 0.7-2.3 hours were obtained with MFOSC-P instrument on PRL 1.2m Mt. Abu telescope, covering \hb and \ha wavelengths. Coupled with the data available in the literature and archival photometric data from TESS and Kepler/K2 archives, various variability parameters are explored for any plausible systematics with respect to their spectral types, and rotation periods. Though about 64\% of our sample shows statistically significant variability, it is not uniform across the spectral type and rotation period. \ha activity strength (L$_{H\alpha}$/L$_{bol}$) is also derived and explored for such distributions.}

\begin{document}

\maketitle

\section{Introduction}
\label{sec-intro}
A significant number of M dwarfs are found to be magnetically active \citep{West2015}, and it is expected that the nature of magnetic activities (coronal mass ejection, flares, strong stellar winds, star spots, etc.) would affect the evolution of its companion. It has also been proposed that these magnetic activities are closely related to the rotating periods of the underlying M dwarfs \citep{West2008, Reiners2012a, Reiners2014, Newton2017, Wright2018}. In comparison to other chromospheric emission lines like Ca II, Mg II, and K lines found in the faint blue region of the spectrum, the chromospheric \ha emission line is commonly utilized for activity-related studies \citep{Walkowicz2009A}.
\par
Magnetic activities in the M dwarf occur on time scales ranging from a few seconds to several hours \citep{Kowalski2010, Yang2017}. However, we notice that there is a scarcity of systematic spectroscopic studies of M dwarfs samples with shorter cadence (5 minutes or less) in the literature. While several studies have been made on a sample of M dwarfs across spectral type \citep{Lee2010, Hilton2010, Kruse2010, Bell2012}, except \cite{Lee2010}, most of these studies had explored the variability in H$\alpha$ at the cadence greater than 15-20 minutes or with a sample of uneven cadence. As a result, shorter-duration behavior could not be investigated, leaving a gap in our systematic understanding of \ha variability on such a temporal scale. Therefore, we conducted a systematic short-term (mostly 5 minutes individual frame exposures spanning 0.7-2.3 hours) spectroscopy monitoring of a sample of M dwarfs in the spectral range of M0-M6.5 to investigate their short-term \ha variability. This spectral range was suitable because our sample of 83 M dwarfs in the M0-M6.5 spectral types complemented the data set studied by \cite{Lee2010} in the M3.5-M8.5 range. As a result, we constructed a sample of 126 sources spanning the whole M0-M8.5 spectral range. The photometric light curves from the TESS and Kepler/K2 missions are used to calculate rotation periods in order to investigate the possible distribution of rotation periods, activity, and spectral types.

\section{Sample selection \& Observations}
\label{sec-ObsMtAbu}

\begin{figure}
\centering
\includegraphics[angle=0,width=0.4\textwidth]{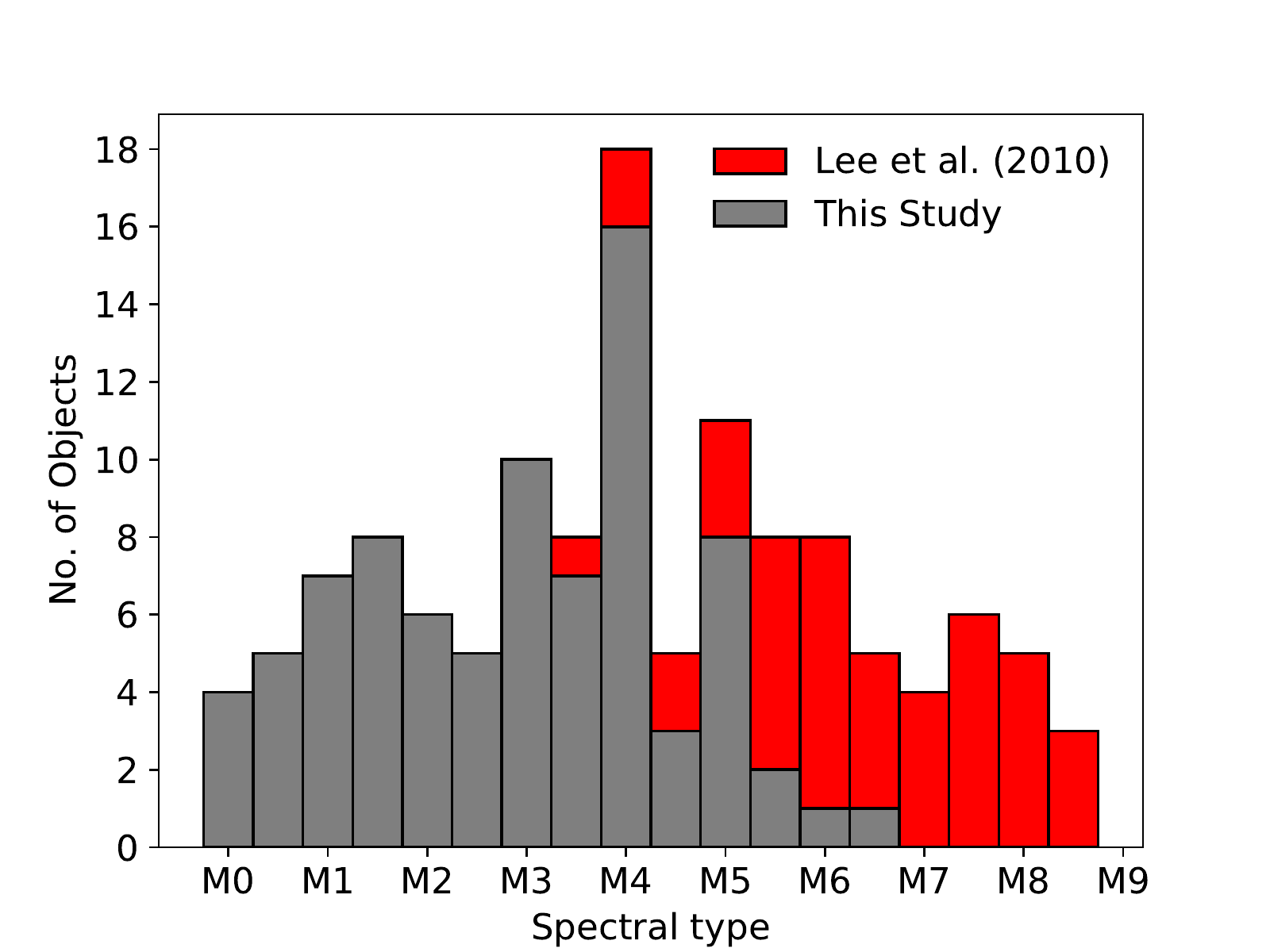}
\caption{Distribution of 83 M dwarfs of this study along with 43 M dwarfs from \protect \cite{Lee2010} with respect to the spectral type.}
\label{figure1}
\end{figure}

We used the MFOSC-P instrument \citep{Srivastava2018, Srivastava2021, Rajpurohit2020} on PRL 1.2 m, f/13 telescope to observe our sample of M dwarfs (spectral region M0-M6.5). Because of the telescope's moderate aperture and combined telescope + instrument efficiency, we often limited our sample to V magnitudes brighter than 14. This also limits us to extend the spectral range beyond M6, as most of the late M dwarfs (M7 and beyond) are too faint (V$>$16) for spectroscopy with MFOSC-P on a 1.2m telescope. However, as discussed in section~\ref{sec-intro}, this spectral range would complement the sample of \cite{Lee2010}. The distribution of all sources is depicted in Fig.~\ref{figure1}. The sources used for this study had typical \ha equivalent widths (EW) of $\sim$ -0.75 \AA~, which corresponds to the detectable \ha line in emission. As a result, we chose 83 appropriate targets from the lists of \cite{Jeffers2018} and \cite{lepine2011}. They were observed from March 2020 to March 2021. We used the R1000 mode of MFOSC-P (dispersion $\sim$1.9 \AA~ per pixel) with a slit-width of 1 arc-second having a spectral range of 4700-6650\AA~. The targets were monitored for $\sim$0.7-2.3 hours in a single stretch with integration times ranging from 200-600s per frame for each of the sources. As a result, each data set typically consists of $\sim$8-18 frames of the single spectrum. Further information on the MFOSC-P instrument and data reduction process can be found in \cite{Srivastava2018, Srivastava2021, Rajpurohit2020}. Photometric light curves of 75 of the above sources were obtained from the TESS and Kepler/K2 archival databases through Mikulski Archive for Space Telescopes (MAST) portal\footnote{https://mast.stsci.edu/portal/Mashup/Clients/Mast/Portal.html}.

\section{Analysis and Results}
\label{sec-spec_result}

\subsection{H$\alpha$ \& H$\beta$ Equivalent widths and their variability}
\label{SubSec-HaHbEW}

To estimate the Equivalent Widths (EWs) of \ha nd \hb emission lines, the spectral wavebands are taken 6557.6-6571.6 $\AA$ and 4855.7-4870.0 $\AA$ respectively (following  \cite{Hilton2010}). The corresponding continuum regions are  6500-6550 $\AA$ \& 6575-6625 $\AA$ for \ha and 4810-4850 $\AA$ \& 4880-4900 $\AA$ for \hb emissions. The average values of the continuum flux in these regions are used for the EWs estimations while summing the area under the line.

%%%%%%%%%%%%%%%%%%%%%%%%%%%%%%
\begin{figure*}
\centering
\includegraphics[angle=0,width=\textwidth]{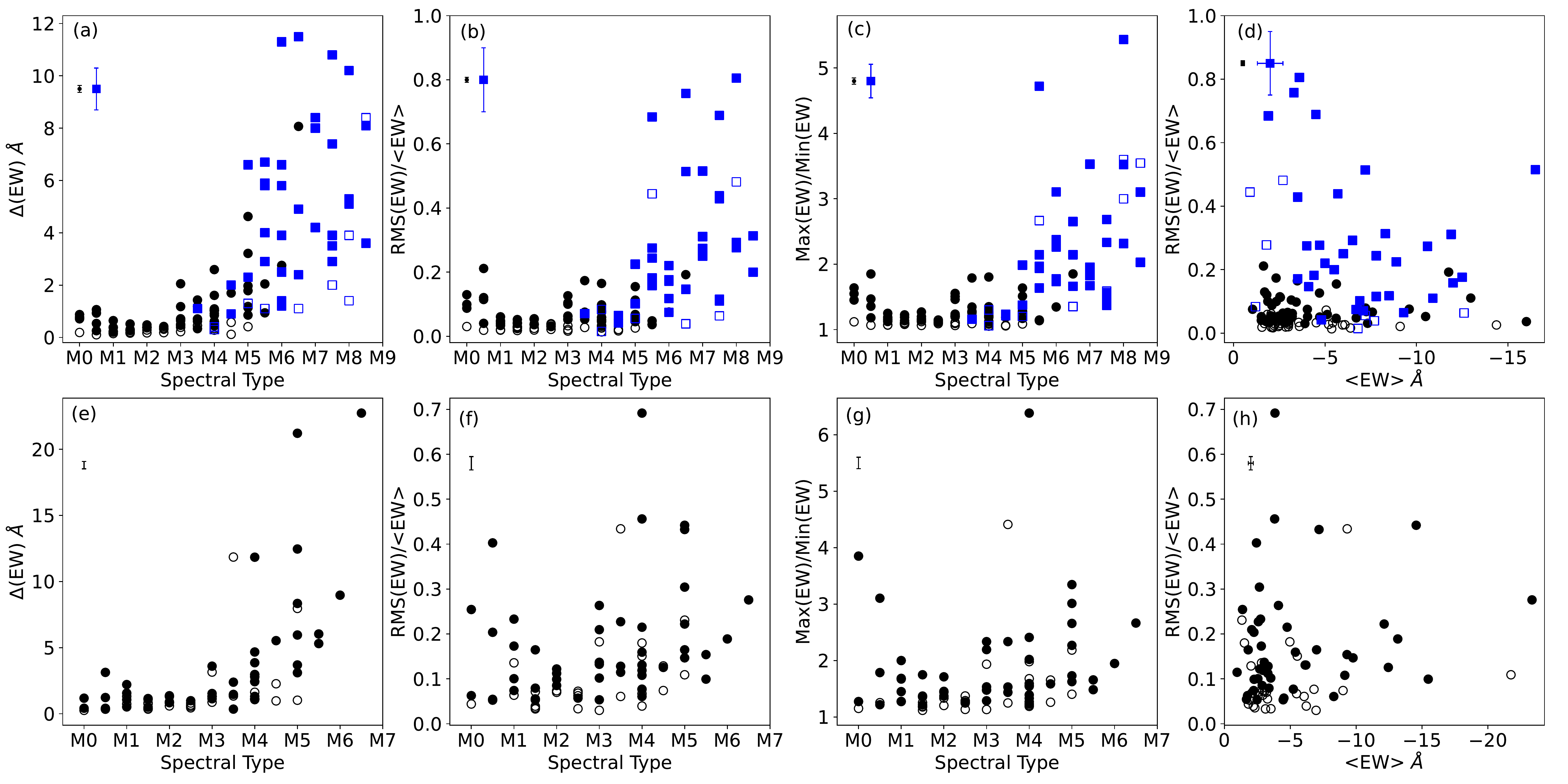}
\caption{ 
Plots showing the variations of various quantities depicting the variability of the EWs of \ha and \hb emissions in M dwarfs. Panels (a), (b), and (c) (in the top row for \ha) and panels (e), (f), and (g) (in the bottom row for \hb) show the changes in  $\Delta$({\rm EW}), ${\rm RMS (EW)}/\langle {\rm EW}\rangle$, and $\rm Max(EW)/ Min(EW)$ respectively as the function of spectral type. Panels (d) and (h) show the variation of ${\rm RMS (EW)}/\langle {\rm EW}\rangle$ with respect to $\langle {\rm EW}\rangle$ for \ha and \hb respectively. Black circles represent the data points of our sample in this study. Blue squares represent the data points derived from the values given in Table 2 of \protect \cite{Lee2010}. Filled and open circles/squares represent the objects identified with varying and non-varying \ha using the $\chi^2$ criterion. The error bars on the top-left corner show the median errors of the data points.}
\label{fig-Var_plot_HaHb}
\end{figure*}
%%%%%%%%%%%%%%%%%%%%%%%%%%%%%

\begin{figure*}
	\centering
	\includegraphics[angle=0,width=0.95\textwidth]{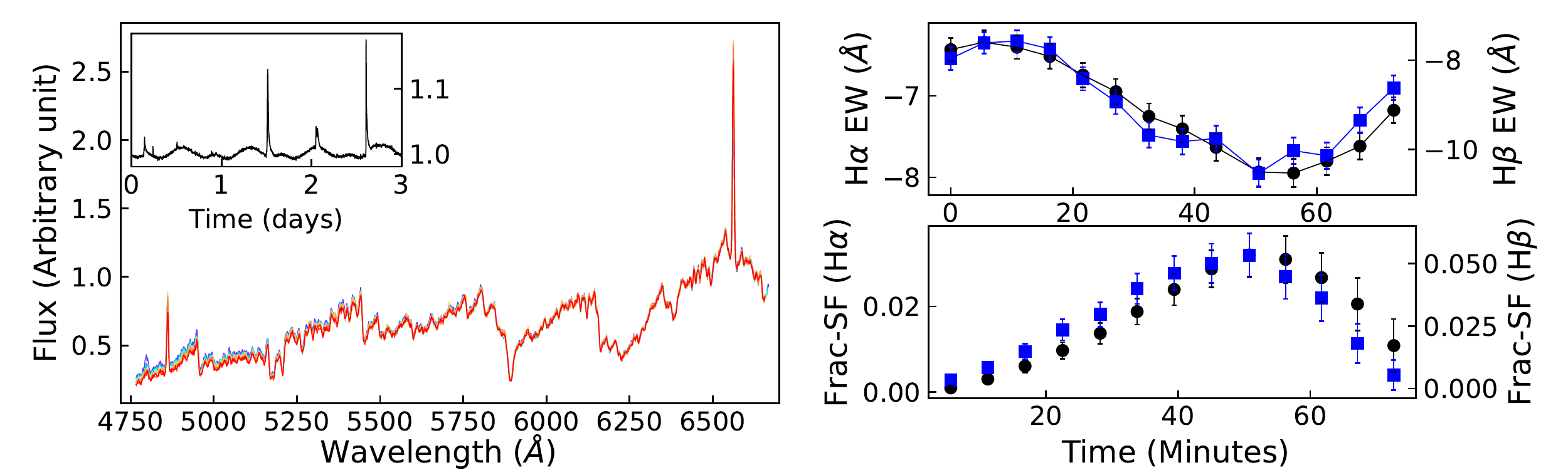}
	
	\caption{The figure shows the time-varying spectra along with their photometric light curves in the inset for one of our sources, PMJ02088+4926 (Spectral type: M4.0, Rotation period: 0.75 days). The corresponding upper and bottom right side panels show the time variations of the EWs of \ha / \hb and fraction structure function (SF), respectively. Data for \ha and \hb are shown in black circles and blue squares, respectively.}
	
	\label{fig-EW-SF}
\end{figure*}
%%%%%%%%%%%%%%%%%%%%%%%%%%%%%%

A $\chi^2$ minimization is done over the EW time series data set (EW light curve) for each of the sources to measure the emission line flux variability in this time series. Calculating $p$-values for given degrees of freedom was used to determine the confidence of the  $\chi^2$ fit. A source is considered variable if its $p$-value is less than 0.05. In our sample of 83 M dwarfs, 30 objects ($\sim36\%$) exhibit no variability in \ha emission with a confidence level of more than $95\%$ ($p$-value $<$0.05). We utilize the metrics$\Delta{\rm EW} = Max(EW) - Min(EW)$,  ${\rm RMS(EW)}/\langle {\rm EW}\rangle$ and $R({\rm EW})= \rm Max(EW)/ Min(EW)$ to measure the variability strength for both the \ha and \hb emission lines. In Fig.~\ref{fig-Var_plot_HaHb}, we see a clearly rising trend, as previously observed by \cite{Lee2010}, and \cite{Kruse2010}, indicating more variability in the later spectral type of M dwarfs. Panels (d) and (h) of Fig.~\ref{fig-Var_plot_HaHb} indicate the segregation of our data set (M0-M6.5) and the \cite{Lee2010} data set (M3.5-M8.5). The later types of M dwarfs, though having a lower  $\langle {\rm EW}\rangle$, are more variable. For our sources (M0-M6.5), the values of ${\rm RMS(EW)}/\langle {\rm EW}\rangle$ are found to be less than $\sim$0.2 for \ha and less than $\sim$0.5 for \hb.

\par
We attempted to investigate the time scales of this variability, by constructing a simple fractional structure function (SF). Though the fractional SFs do not clearly demonstrate any time scale of variability, they do corroborate an interesting trend seen by \cite{Bell2012}. The sources observed to vary over a longer time period showed a fractional SF with an increasing trend, as expected. However, the sources whose EWs light curves are seen to vary at shorter time scales show a roughly flat distribution of fractional SF at all times, as also demonstrated by the results of \cite{Bell2012}. \cite{Bell2012} attributed such high variable source behavior to a variability time scale shorter than their shortest time-separation bin of $\sim$15 minutes. Even at a cadence o f$\sim$5 minutes, we see the same tendency. These findings are discussed in section~\ref{Sec-Discussion} along with other results. For one of the sources, the spectra, EW light curves, and fractional SFs for \ha and \hb are shown in  Fig.~\ref{fig-EW-SF}.

\subsection{\ha and \hb activity strength}
\label{SubSec-HaHb-Strength}

\begin{figure*}
	\centering
	\includegraphics[angle=0,width=0.7\textwidth]{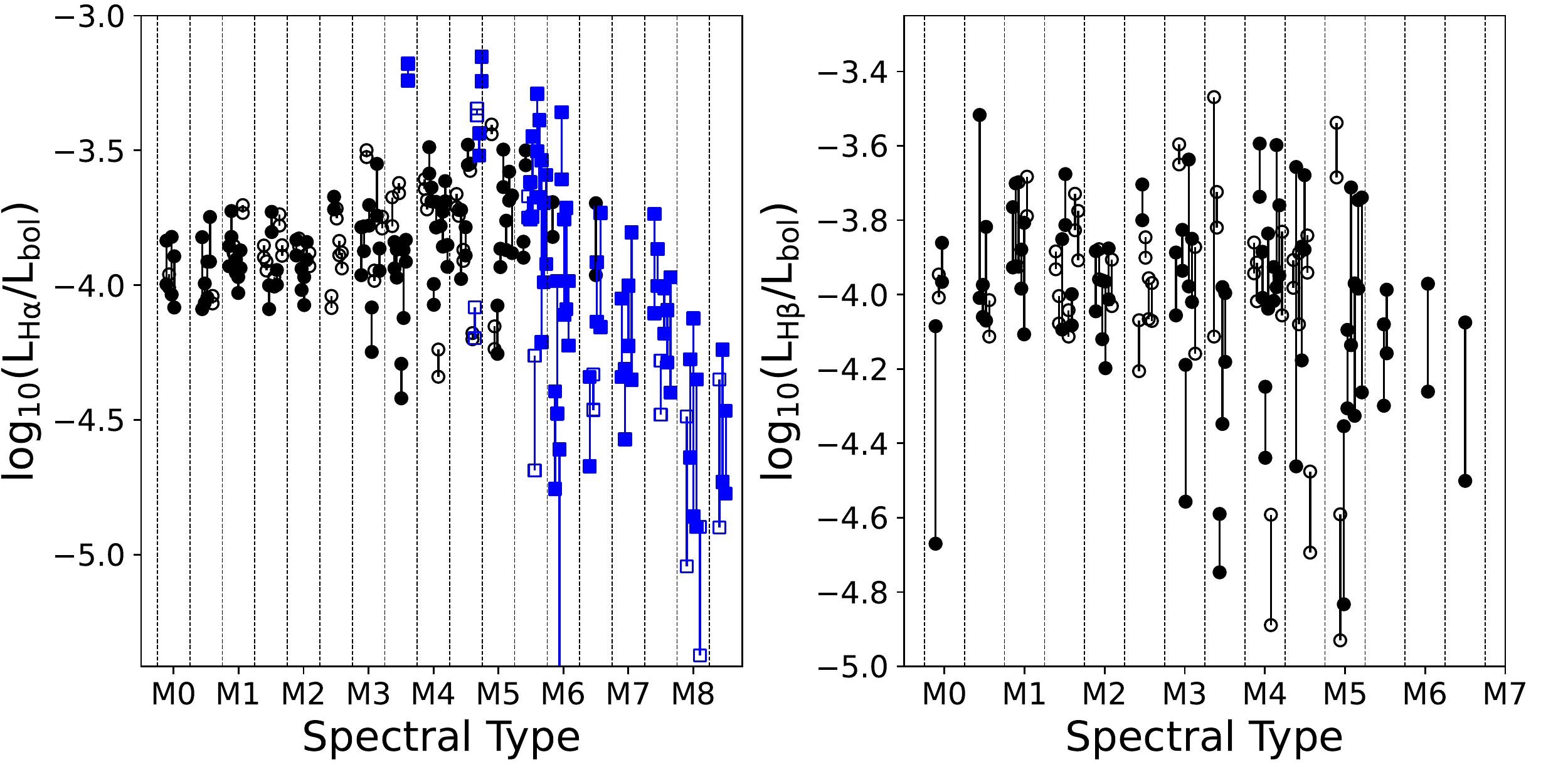}
	\caption{The distribution of derived activity strengths ($L_{\rm H\alpha}/L_{\rm bol}$ and $L_{\rm H\beta}/L_{\rm bol}$) for \ha and \hb (left panel for \ha and right panel for \hb) with respect to their spectral type. The solid lines connect the maximum and minimum activity strength values measured for each source. The positions of objects are displaced horizontally within vertical solid lines for clarity. Symbols have the same meaning as in Fig.~\ref{fig-Var_plot_HaHb}.}
	\label{fig-Strength_HaHb} 
\end{figure*}

\ha or \hb activity strength is defined as the ratio of their luminosity to the bolometric luminosity \citep{West2008, Hilton2010, Lee2010, Newton2017}. The activity strength represents a more meaningful estimation of activity between stars of various masses than EW alone \citep{Reid1995b} since it indicates the relevance of the line flux relative to the star's total energy output. To compute the activity strength for the \ha and \hb emission lines, we used the relationships provided by \cite{Douglas2014}. The chi factor, which is required to calculate the activity strength, is derived from photometric color $(i-J)$ \citep{Walkowicz2004, Douglas2014, West2008b}. The derived values of $L_{\rm H\alpha}/L_{\rm bol}$ and $L_{\rm H\beta}/L_{\rm bol}$ are shown in Fig.~\ref{fig-Strength_HaHb} with respect to their spectral type. The values from Table 2 of \cite{Lee2010} are also included in the \ha plots.  L$_{H\alpha}$/L$_{bol}$ reaches a constant value of $\sim$10$^{-3.8}$ for M dwarfs with spectral types M0-M4, then decreases for later spectral types (later than M4), indicating decreased activity strengths for the later types M dwarfs. However, the variability is higher for these later spectral types, which can be approximated as the ratio of maximum to minimum values (of the \ha and \hb line flux for a given time series of an M dwarf). This indicates that, while the level of activity in these later type M dwarfs is low, they are more variable. These findings are discussed in section~\ref{Sec-Discussion} along with other results.

%%%%%%%%%%%%%%%%%%%%%%%%%%%%%%

%%%%%%%%%%%%%%%%%%%%%%%%%%%%%%%%

\subsection{Rotation Period}
\label{SubSec-RotationPeriod}

\begin{figure*}
	\centering
	\includegraphics[angle=0,width=\textwidth]{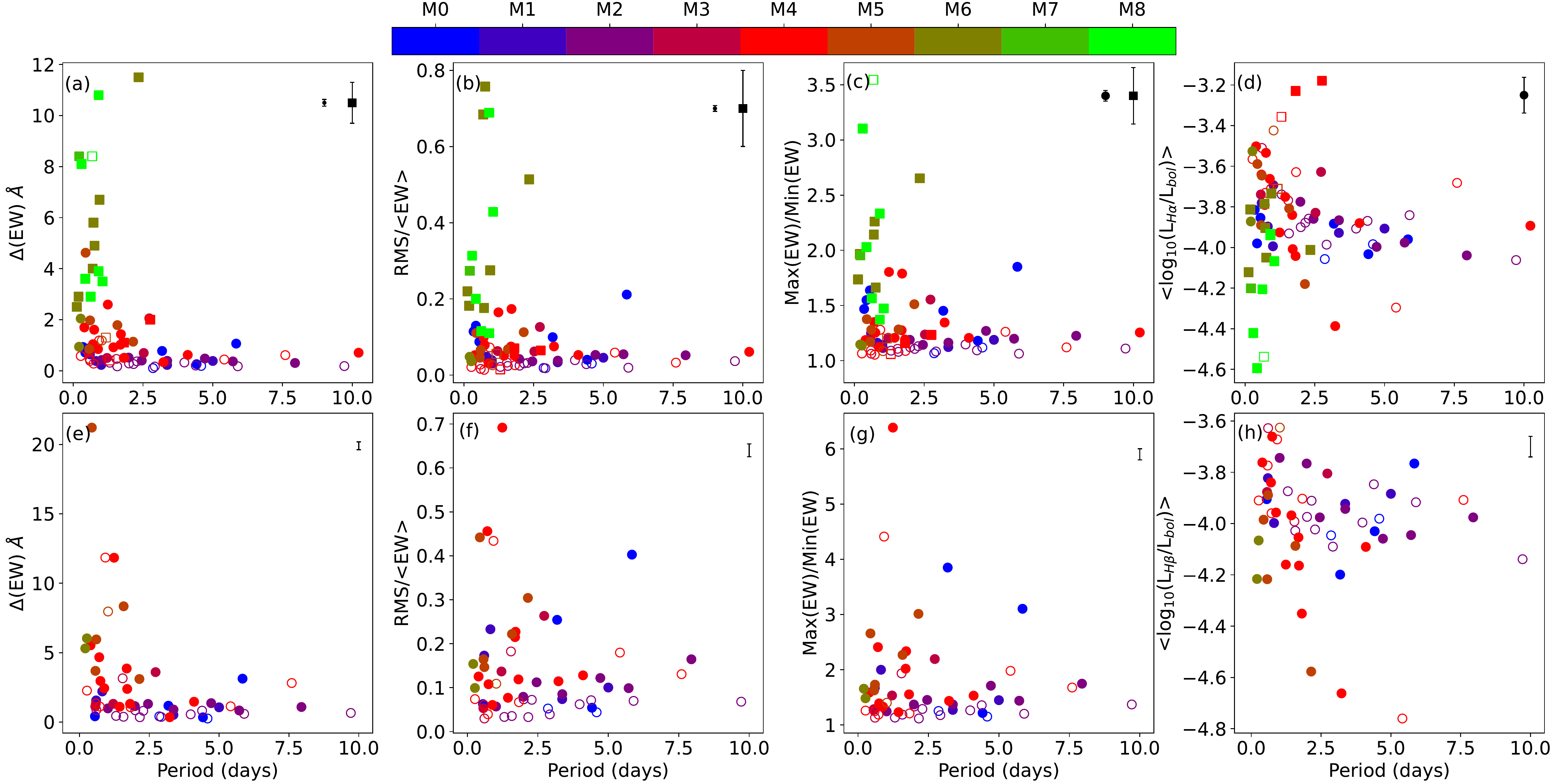}
	\caption{Figure shows the distribution of $\Delta({\rm EW})$, ${\rm RMS(EW)}/\langle {\rm EW}\rangle$, $R({\rm EW})= \rm Max(EW)/ Min(EW)$ and mean activity strengths with their rotational periods. The top panels are for \ha and the bottom panels are for \hb emission. The median error bars of the data points are shown in the top-right corner. Symbols have the same meaning as in Fig.~\ref{fig-Var_plot_HaHb}.}
	\label{fig-variability-wrt-period-HaHb}
\end{figure*}

TESS and Kepler/K2 missions \citep{Caldwell2010, Koch2010, Howell2014, Ricker2015} have delivered excellent cadence and photometric precisions for a wide range of stars in the last decade. These photometric light curves were used to calculate the rotation periods of the objects in our sample. The rotation periods, P$_{rot}$, are calculated by quantifying the periodic brightness oscillations in the light curve induced by starspots on the objects' surfaces. These were determined using a periodogram technique such as the Lomb-Scargle periodogram \citep{Lomb1976, Scargle1982}. The rotation periods of 82 objects were estimated using the above method from 106 sources where light curves from TESS and Kepler/K2 were available. They are found to be in the $\sim$0.2-10 day range. Fig.~\ref{fig-variability-wrt-period-HaHb} depicts the distribution of the variability indicators with respect to rotation periods.

\par
In the past, many magnetic activity indicators were utilized to investigate the relationship between magnetic field strength and star rotation \citep{Douglas2014, West2008, West2015, Newton2017, Jeffers2018}. Similar to \cite{West2015} and \cite{Jeffers2018}, we find that M dwarfs with longer periods exhibit less variability. It is also known that the strength of activity in M dwarfs decreases with increasing rotation period \citep{West2015, Jeffers2018}. When we consider the spectral types, we can see an interesting behavior in the panels $a$, $b$, $c$ (for \ha) and $e$, $f$, and $g$ (for \hb). The short-term variability indicators show a significant increase in variability for the faster rotating M dwarfs having periods of $<$ 2 days and the majority of these objects belong to the later types (M4-M8). Because these later types are often fast rotators, such significant variability could very likely be caused by the magnetic field coupling with the rotation-induced dynamics of chromospheric regions. The magnetic properties of stars are known to be connected to \ha emissions and star-spots \citep{Newton2017}. As a result, such apparent correlations may be plausible and expected.

%%%%%%%%%%%%%%%%%%%%%%%%%%%

\section{Discussion}
\label{Sec-Discussion}

The EW light curves of \ha and \hb emissions exhibit a significant gradient in variability amplitudes over a time scale of a few minutes to a few tens of minutes. The derived mean activity strengths ($\langle$L$_{H\alpha}$/L$_{bol}$$\rangle$ and $\langle$L$_{H\beta}$/L$_{bol}$$\rangle$) across this short term time series agree with the trend seen at other time scales \citep{Bell2012, Lee2010}, where the activity strengths decrease for later spectral types and the corresponding variability increases. The activity strengths($\langle$log$_{10}$(L$_{H\alpha}$/L$_{bol}$)$\rangle$) in our dataset of 126 sources are close to $\sim$-3.8 for the spectral types M0-M4 and then decline to $\sim$-5.0 for mid to late M dwarfs. This is very similar to the trend seen in \cite{Kruse2010} and \cite{Bell2012} for a larger cadence. These activity strength breaks could be explained by a change in the magnetic dynamo process at the fully convective boundary. While early M dwarfs possess a partially convective envelope, late M dwarfs (M4 and up) have a fully convective envelope \citep{Reiners2012a, Reiners2014, Newton2017, Wright2018}.

\par
In this work, the derived rotation periods of M dwarfs range between $\sim$0.2-10 days. Higher variability was seen for stars with rotation periods of $<$ $\sim$2 days. The majority of these are late-type M dwarfs (M6-M8.5). This behavior could be explained by the magnetic heating of the stellar atmosphere, which is a result of the underlying magnetic dynamo's significant dependency on stellar rotation. In the literature, such behavior has been thoroughly investigated\citep{West2015, Newton2017, Wright2018, Raetz2020}. As a result, the observed behavior in the \ha variability with respect to the rotation period is completely plausible.

\par
High-activity objects are found to be less variable, possibly because more energetic events (such as large flares) are required to change their observational status in terms of \ha / \hb strengths. Such intense events may necessitate longer time scales to originate and/or evolve. Low energy events that might occur on shorter time scales would most likely govern the variability of less active stars. Thus, studies of the time scales of the \ha / \hb variations are crucial to understanding the underlying activity on the star's surface. A larger sample with finer time resolution could be the key to understanding the physical mechanisms behind such behaviors.

\section*{Acknowledgments}
{The research work at the Physical Research Laboratory is funded by the Department of Space, Government of India. We thank Veeresh Singh (PRL) and Rishikesh Sharma (PRL) for their useful discussions. This research has made use of the SIMBAD database, operated at CDS, Strasbourg, France. This paper includes data collected by the TESS and Kepler mission and acquired from the Mikulski Archive for Space Telescopes (MAST) data archive at the Space Telescope Science Institute (STScI). Funding for the TESS and KEPLER missions is provided by NASA's Science Mission Directorate. STScI is operated by the Association of Universities for Research in Astronomy, Inc., under NASA contract NAS5-26555.}

\bibliographystyle{cs21proc}
\bibliography{example.bib}

\end{document}